\documentstyle[aps,prb,multicol,epsf]{revtex} 
\begin{document}   
\draft 
\title{The leading Ruelle resonances of chaotic maps}     
                   
\author{Galya Blum$^1$ and Oded Agam$^2$}  
  
\address{$^1$ Department of Condensed Matter Physics, Weizmann Institute 
of Science, 76100 Rehovot, Israel \\
$^2$ The Racah Institute of Physics, The Hebrew University, Jerusalem 
91904 Israel} 
\maketitle 
\begin{abstract} 
The leading Ruelle resonances of typical chaotic maps,
the perturbed cat map and the standard map,
are calculated by variation. It is found that, excluding the 
resonance associated with the invariant density, the next 
subleading resonances are, approximately, the roots of the  equation
$z^4=\gamma$, where $\gamma$ is a positive number
which characterizes the amount of stochasticity
of the map. The results are verified by numerical computations, and
the implications to the form factor of the corresponding
quantum maps are discussed. 
\end{abstract}      
\begin{multicols}{2} 
The statistical charectaristics of the quantum mechanical
spectrum of a system, which has a classical analog, are
related to its underlying classical dynamics. 
For example, in chaotic systems, this relation  
is revealed by Gutzwiller's trace formula\cite{Gutzwiller} 
which expresses the  density of states in terms of a sum over  
the classical periodic orbits of the system. In disordered systems,
spectral properties of the quantum system are expressed in terms
of the diffusons and the Cooperons diagrams associated with the classical  
diffusive modes of the system\cite{Altshuler85}. 
 
There is, however, a difference between the periodic orbit  
picture and the diagrammatical approach of disordered systems. The first 
uses the individual trajectories of the systems as the basic ingredients 
for the semiclassical expansions, whereas the second approach employs
spectral properties of the evolution operator 
of classical distribution functions. The latter approach is, 
therefore, more suitable for a field theoretic treatment. 

Recently an extension of the field theoretic formalism, from disordered 
systems to general chaotic ones, has been proposed\cite{NLSM}. The analog
of classical diffusion, in the generalized case, is the coarse grained
Liouville dynamics of distribution functions\cite{AAA} in the limit
of zero coarse graining\cite{AL}. The corresponding evolution operator 
is the Frobenius-Perron (FP) operator, and its eigenvalues are the 
Ruelle resonances\cite{Ruelle86}.
These eigenvalues describe the decay of classical
correlation functions of smooth
observables.
   
The association of spectral properties of quantum chaotic systems
with their classical spectra calls for a study of the
Ruelle resonances of general 
chaotic systems. The central question is whether chaotic systems can be
classified into equivalence classes according to their classical spectrum? 
An obvious example of such a class
is systems where the effective dynamics is  diffusive. For instance,
the kicked rotor map exhibits diffusion in momentum space\cite{Chirikov79},
and the stadium billiard, in the limit of small
distance between the semicircles, is diffusive in the
angular momentum space\cite{Borgonovi96}.  

Are there other universality classes? In principle, 
there should be. Diffusion is not the only type of universal 
dynamics of chaotic systems where correlations are rapidly lost. 
Other possibilities can be, for instance,  various types of Levy 
flights\cite{Mandelbrot82}.

Yet, the lack of approximation schemes 
for calculating Ruelle resonances, obstructs the
characterization of the effective dynamics of
general  chaotic systems. The physical approach for
calculating  Ruelle resonances 
is to project the dynamics, from the  full phase space, down to the 
manifold on which the dynamics is slow, and
to construct an equation for the probability density on this manifold.
For example, in disordered systems, the fast and the slow manifolds are
the momentum and the real space respectively.
Projecting the dynamics down to real 
space yields the diffusion equation, whose eigenvalues constitute the 
leading Ruelle resonances of the system.
In trying to apply a similar procedure for
general chaotic systems, one encounters the problem  
of identifying  the slow and the fast manifolds. These
can be complicated functions in phase space.
 
In this paper we choose a different approach.
We will construct a simple variational method for calculating
the leading resonances, and apply it to two generic maps: 
The perturbed cat map\cite{Matos95} and the 
standard map\cite{Chirikov79}. The results 
indicate that, indeed, the leading Ruelle resonances of these maps
share some universal features (different from diffusion). 
Namely, the configuration of the leading resonances 
in the complex plain is similar. Excluding the 
eigenvalue associated with the invariant density, $z_0=1$, the next 
subleading resonances are, approximately, the roots of the 
equation: $z^4=\gamma$, where 
$\gamma$ is a real positive number, smaller than unity, which decreases as
the map becomes more stochastic. It will be shown that this 
configuration of resonances is significant for the
spectral statistics of the corresponding quantum maps. For example,
in the form factor,  it leads to
a suppression of the nonuniversal corrections to the results 
of random matrix theory\cite{Mehta91} (RMT).

In introducing our ideas it will be instructive to consider a specific
example. We begin with the perturbed cat map\cite{Matos95} which is 
an area preserving map from the unit torus to itself: 
\begin{eqnarray} 
\begin{array}{l} x_{n+1} = 2 x_n + y_n \\ 
                 y_{n+1}= 3 x_n + 2 y_n +  f(x_{n+1}) 
\end{array} ~~~(\mbox{Mod}~1). \label{map} 
\end{eqnarray} 
Here $(x_n,y_n)$ are the phase space coordinates at discrete  

{\narrowtext    
\begin{figure}      
  \begin{center}     
\leavevmode     
        \epsfxsize=4.4cm        
         \epsfbox{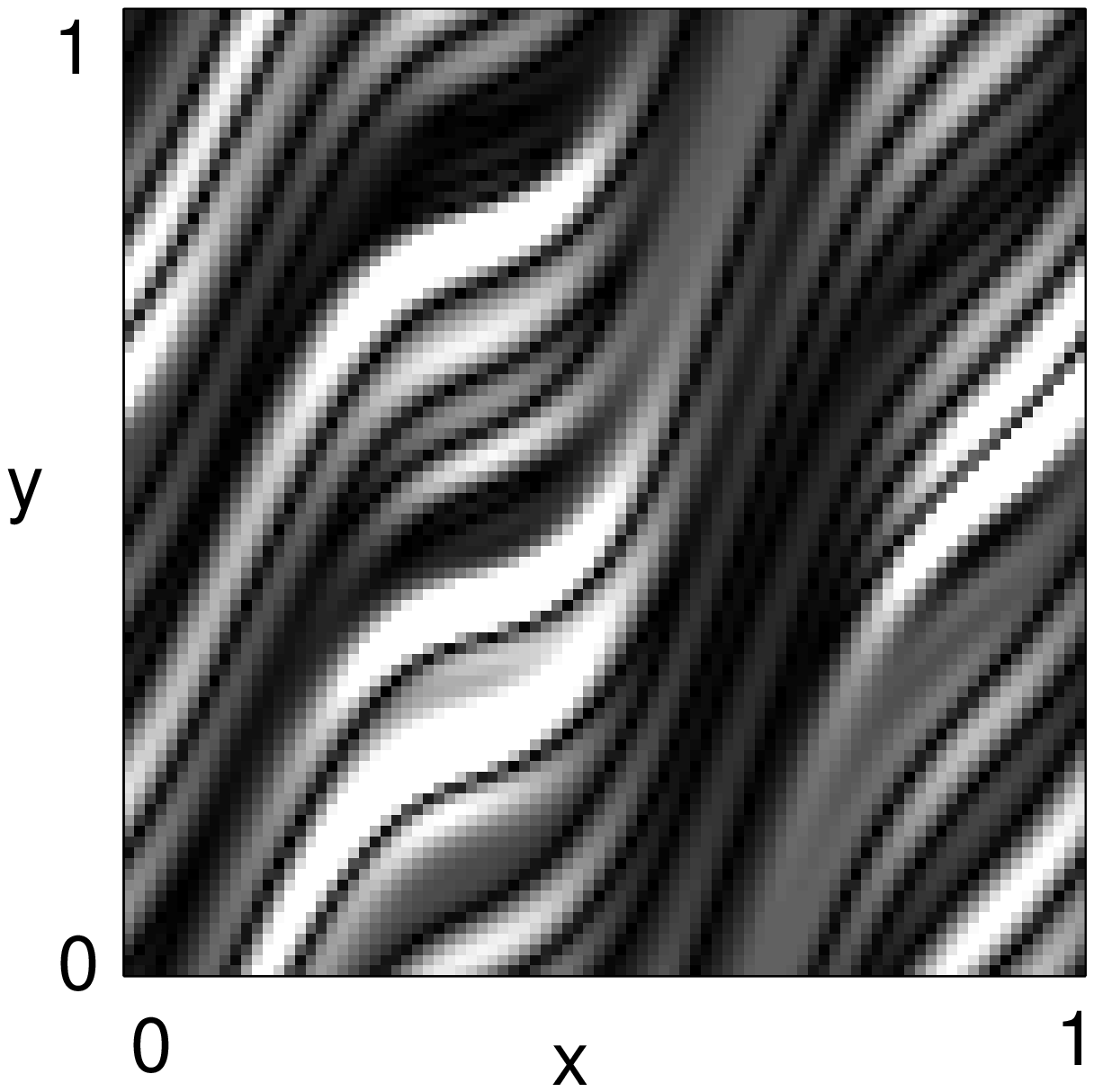}      
\leavevmode     
        \epsfxsize=4.3cm        
         \epsfbox{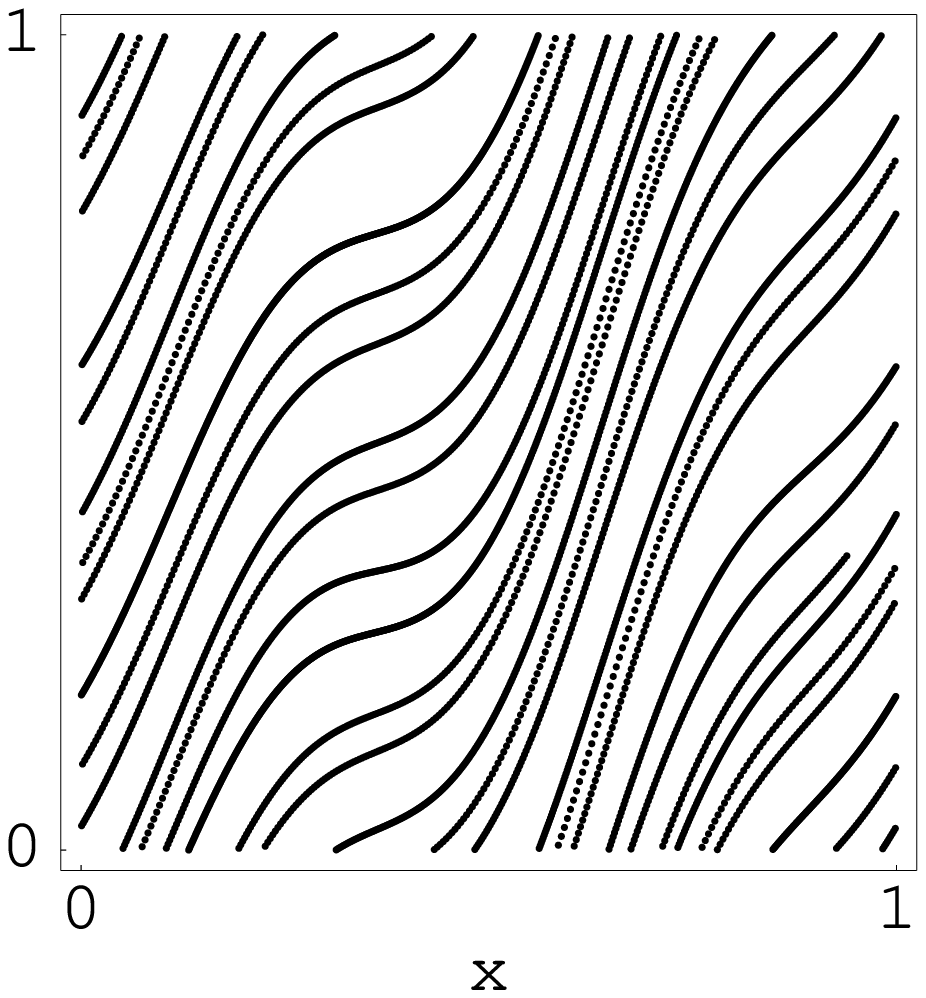}   
\end{center}     
   
	\caption{ Left panel: The absolute value of the  
right eigenfunction, $\psi_1^R$ of the perturbed cat map at $k=0.5$ and 
coarse graining $s=0.001$. The dark and the bright regions correspond to 
the minima and maxima  of $|\psi_1^R|$, respectively. This function
follows the unstable manifold of the map which is
shown on the right panel.
}               
\end{figure}        
}
\noindent 
time $n$, and 
\[ f(x)= \frac{k}{2 \pi} \left[ \cos ( 2 \pi x) -  
\cos ( 4 \pi x) \right] \] 
is a perturbation function. 
This function is smooth and periodic on the torus, therefore, 
the map is continuous, and differentiable.
At $k=0$ the map, known as Arnold's cat map\cite{Arnold68}, 
is highly chaotic\cite{Antoniou97}. 
Namely, time correlation functions decay faster than 
exponential. This behavior 
is related to the presence of hidden symmetries of
number theoretical origin\cite{keating91}.
The role of $f(x)$ is to break these symmetries. As a result, 
at finite values of $k$, relaxation is exponential. It has 
been also proved that for $k<0.11$ all the fixed points of (\ref{map}) are 
in one to one correspondence with the original cat map. 
Consequently, all orbits are  unstable and the system 
can be characterized as ``hard chaotic''\cite{Matos95}.

The classical propagator of the map, for one time step, is  
\begin{eqnarray} 
\langle x',y'|U| x,y \rangle = \delta_p[x'\!-\! 2x \!-\!y] \delta_p[ y'\!-\! 
3x \!-\! 2 y \!-\! f(x' )], \label{U} 
\end{eqnarray} 
where $\delta_p(x)$ is a periodic $\delta$-function on the torus. Due to 
Liouville's theorem, the above propagator is unitary when acting on
Hilbert space of square integrable functions. Within this space
the dynamics is reversible, however, as soon as the 
dynamics is coarse grained, it becomes irreversible and
phase space densities relax to the invariant density. 
 
A simple way of coarse graining the
dynamics is to give the $\delta$-functions in (\ref{U}) a  
finite width, for instance, by the replacement 
\[ 
\delta_p(x) \to \sum_n \frac{ 1}{\sqrt{\pi s}}e^{-\frac{ (x-n)^2}{s}},
\] 
where $s$ is the coarse graining parameter.  
The Ruelle resonances of the map
are obtained by diagonalizing the propagator at a finite value of $s$, 
and taking $s$ to zero {\em after} the diagonalization. It  can be shown
that this procedure is  
equivalent to the introduction of noise into the equations of motion  
(\ref{map}), calculation of the eigenvalues  
averaged over the noise, and finally taking the limit of noise to 
zero\cite{Gaspard95}. The resulting resonances 
have absolute values smaller than unity, 
except for the resonance associates with the invariant density, $z_0=1$. 

Since the coarse grained evolution operator is not unitary, 
the left eigenfunctions of $U$ are not the complex 
conjugates of the right eigenfunctions. Instead they can be viewed as
the right eigenfunctions of the propagator associated with the inverse map
(i.e.~the propagator taking the system one time step backward).
Thus the eigenvalue equations are: 
\begin{eqnarray} 
U|\psi^R_\alpha \rangle = z_\alpha  |\psi^R_\alpha \rangle, 
~~~~~~\langle \psi^L_\alpha | U = \langle \psi^L_\alpha | z_\alpha, 
\label{diag} 
\end{eqnarray} 
where $|\psi^R_\alpha \rangle$ and $\langle \psi^L_\alpha |$ are the right  
and the left eigenfunctions associated with the resonance $z_\alpha$.
According to Ruelle's theorem\cite{Ruelle86}, these eigenfunctions become
distributions in the limit of zero coarse graining. The distributions
associated with the left and the right eigenfunctions
follow the stable and the unstable manifolds respectively.
Nevertheless, they satisfy the normalization condition 
\begin{eqnarray}  
\langle \psi^L_\alpha |\psi^R_\beta \rangle=\delta_{\alpha,\beta}. \nonumber
\end{eqnarray} 

In the left panel of Fig.~1 we show an example of the right eigenfunction,
$\psi_1^R$, of the perturbed cat map for $k=0.5$ and $s=0.001$.
As can be seen, this eigenfunction follows the unstable manifold 
depicted on the right panel. As the coarse graining parameter, $s$, 
reduces, the eigenfunction becomes more singular, in accordance with 
Ruelle's theorem.

In what follows we shall calculate the leading Ruelle resonances
of the map (\ref{map}) using variational approach which takes 
into account the structure of the eigenfunctions illustrated 
in Fig.~1. For this purpose we 
construct the functional 
\begin{eqnarray} 
F= \langle L | U | R \rangle - z  \langle L | R \rangle, \label{F} 
\end{eqnarray} 
where $|R \rangle$ and $\langle L|$ are the right and the corresponding
left trail functions. The leading Ruelle resonances are obtained by 
variation\cite{comment1} of $F$ with respect to  $\langle L |$. 

The success of a variational calculation depends on the extent to which the
variational wave functions capture the physics of the problem.  
In our case, one should construct  left and  right trail functions which 
follow the stable and the unstable manifolds respectively. We will
use the dynamics itself to generate these functions
by repeated applications of the map on a smooth initial state. 
Thus, our right trail function takes the form
\begin{eqnarray} 
| R\rangle = \sum_{ n= -\infty}^{\infty}\!\!\!\!\!\!\ ^\prime
 A_n |n \rangle, 
~~\mbox{with}~~|n \rangle = e^{ 2 \pi i \phi_n^R(x,y)}, \label{Rvar} 
\end{eqnarray} 
where $A_n$ are amplitudes to be determined by variation, 
and the phases $\phi_n^R(x,y)$ are defined by the recursion equation:
\[ 
\phi_n^R(x,y)= \phi^R_{n-1}(2x -y+ f(x), 2y - 3 x - 2 f(x)) 
\] 
with the initial phase $\phi^R_1(x,y) = x$. We define 
$\phi^R_{-n}$ to be $ -\phi^{R}_n$, and the prime 
($\ ^\prime$) indicates that the sum does not include the $n=0$ term. 

The above sequence of phases, where the first few 
of them are
\begin{equation} \begin{array}{l} 
\phi^R_1= x, \\ 
\phi^R_2= 2x -y + f(x), \\ 
\phi^R_3 = 7x -4y + 4f(x) + f(\phi^R_2), \\ 
\phi^R_4 = 26 x - 15 y + 15 f(x) + 4 f(\phi^R_2) + f(\phi^R_3),
\end{array} \label{Rphases}
\end{equation}
is constructed by application of the map (\ref{map})
on the initial smooth state, $e^{i 2 \pi x}$. Therefore,  
$|2 \rangle, |3 \rangle, \cdots$ become increasingly
smooth along the unstable manifold and singular along the stable one.

In the same manner we construct the left function by repeated application 
of the inverse map, namely  
\begin{eqnarray} 
\langle L | = \sum_{ m= -\infty}^{\infty} \!\!\!\!\!\!\!\ ^\prime A^*_m \langle m |~~~\mbox{with}~~
\langle m | = e^{- 2 \pi i \phi_m^L(x,y)},
\label{Lvar} 
\end{eqnarray} 

\noindent
where the left phases are defined by the recursion relation 
\[ 
\phi_m^L(x,y)= \phi_{m-1}^L(2x +y , 3 x+2 y + f(2x+y)) 
\] 
and $\phi_1^L(x,y)=x$. 
 
There are two simplifying features characterizing
 $|R \rangle$ and $\langle L|$.
First, the operation of the propagator, $U$, on these functions
is simple since, by construction,
\begin{eqnarray} 
U|n \rangle= |n+\mbox{sign}(n) \rangle,~\mbox{and}~~\langle m | U = \langle m+ 
\mbox{sign}(m) |. \label{dnm}  
 \end{eqnarray} 
The second feature is that  
$\langle L |$ and $|R \rangle$ are orthogonal
to a constant function. This way we eliminate the leading Ruelle resonance, 
$z_0=1$, from our problem, since the invariant distribution 
is constant in phase space.
The proof for this orthogonality is  straightforward. Let 
$|0 \rangle \!=\!\langle 0 |\!=\!1$ denote the invariant distribution. Then
$\langle 0| n \rangle = \langle 0|U^{n-1}| 1 \rangle = 
\langle 0| 1 \rangle = 0$, where we have used the definition  
of the invariant distribution: $\langle 0|U =\langle 0|$, and that
$|1 \rangle = e^{2 \pi ix}$ is orthogonal to a constant. 
Now, from (\ref{Rvar}) it immediately follows that 
$\langle 0| R \rangle =0$. Similarly it is straightforward to
prove that  also $\langle L| 0 \rangle=0$.

Variation of $F$, given 
by (\ref{F}), (\ref{Rvar}) and (\ref{Lvar}), 
with respect to  $A_m^*$ yields the eigenvalue equation: 
\begin{equation} \mbox{Det} \left[\langle m |  
n+ \mbox{sign}(n) \rangle - z \langle m | n \rangle   \right] = 0. \label{EVE} 
\end{equation} 
As it stands,  this equation is not  simpler to solve  than 
the original problem. However, it can be considerably simplified if the  
matrices $\langle m |n \rangle$ and $\langle m |n +\mbox{sign}(n) \rangle$
can be truncated to a small size. As we show below this is 
indeed the situation. 
For this purpose, it is instructive to notice few properties 
of the matrix elements $\langle m | n \rangle$.  The first one is that 
\begin{equation}  
\langle m | n \rangle = \langle m-\mbox{sign}(m) | n +\mbox{sign}(n)  
\label{1prop}
\rangle, \end{equation}
where $|m| > 1$. This is an immediate consequence of  
$\langle m | n \rangle =\langle m | U^{-1}U  n\rangle$, and the relations 
(\ref{dnm}). The second property of the matrix elements
comes from the definition of the phases with negative indices, 
$\phi^{L,R}_{-n} =-\phi^{L,R}_n$:
\begin{equation}
\langle -m | -n \rangle =\langle m | n \rangle^*,~~\mbox{and}~~~ 
\langle m | -n \rangle =  \langle -m | n \rangle^*. \label{2prop}
\end{equation}  
From (\ref{1prop}) it follows that all matrix  elements
$\langle
\pm m | \pm n \rangle$ can be reduced to integrals of the form 
\[
\langle \pm 1 
| \pm l \rangle =\int dx dy e^{  2 \pi i [ \mp x \pm \phi_l^R(x,y)]}, 
\] 
where $l= |m|+|n|-1$.
Notice that the above integrals decrease  rapidly with $l$.
This follows from the nature of the phases $\phi_l^R(x,y)$. Namely,
as $l$ increases, the phases acquire a stronger dependence 
on $x$ and $y$, see e.g.~(\ref{Rphases}).  Thus $\langle m|n\rangle$ is small
for large $|n|+|m|$, and the significant part
of the eigenvalue equation (\ref{EVE}) is associated only with a
small submatrix.
 
With this observation, we turn now to calculate the main
matrix elements of Eq.~(\ref{EVE}).  
First, it is straightforward to see that 
\begin{equation}
\begin{array}{c}
\langle 1|1\rangle =1, ~~\mbox{and} \\
\langle 1 | -1 \rangle  
=\langle 1 | \pm 2 \rangle = \langle 1 |  3 \rangle =  
\langle 2 |  2 \rangle =0.
\end{array} \label{ZME}
\end{equation}
Other matrix elements are given by 
\begin{eqnarray} \begin{array}{l} 
\langle 1 |  -3 \rangle = \langle 2 |  -2 \rangle = T_{4} = 
-\frac{k^2}{8}- i\frac{k^3}{16} + O(k^4), \\ 
\langle 2 | -3 \rangle = T^2_{3} =\frac{k^4}{3072}+ O(k^5), \\ 
\langle 2 | 3 \rangle = |T_{5}|^2 = \frac{k^6}{256}+ O(k^8), \end{array}  
\end{eqnarray} 
where 
\[ 
T_{\nu}= \int_0^1 dx e^{ -2 \pi i[ \nu x + f(x)]}. \] 
The above results demonstrate 
the strong dependence of the matrix elements $\langle m|n\rangle$  
on $|n|+|m|$, when $k<1$.

{\narrowtext    
\begin{table} 
\vskip 0.5cm
\begin{tabular}{|l|l|l|l|l|} 
$k$ & $z_1$ & $z_2$ & $z_{3,4}$ & $z_{5,6}$  \\ \hline 
0.1 & 0.035 & $-0.037$ & $-0.003\! \pm\! 0.050i$ &
$0.002\pm0.034i$  \\ 
& 0.034 &  $-0.036$  & $-0.001\pm 0.035 i$ &  \\ \hline
0.2 &  0.070 & $-0.080$ & $-0.009 \!\pm\! 0.098i$ &
$0.011\pm 0.064i$ \\ 
& 0.066 &  $-0.076$ & $-0.005 \!\pm \! 0.071i$  & \\ \hline

0.3 & 0.109 &  $-0.134$ & $-0.015 \!\pm \! 0.144i$ &
$0.030\pm0.088i$ \\ 

 & 0.094 &  $-0.12$ & $-0.01 \!\pm \! 0.11i$ & \\ \hline

0.4 & 0.153 & $-0.196$ & $-0.018 \!\pm\! 0.188i$  &
$0.066 \pm 0.113i$\\

& 0.12 & $-0.16$ & $-0.02 \!\pm \! 0.14i$ & \\ \hline

0.5 & 0.208 &  $-0.266$ & $-0.023 \!\pm \! 0.229i$ &
$0.111\pm 0.152i$ \\ & 0.14 & $- 0.21$ & $-0.03 \!\pm \! 0.18i$ &
\end{tabular}  
\vspace{0.3cm} 
\caption{The leading Ruelle resonances of the perturbed 
cat map at various values of the perturbation parameter, $k$.
In each case, the top line is the exact numerical value, while 
the bottom line is the result of the  variational calculation (\ref{FR}). 
The eigenvalue, $z_0=1$, associated with the 
invariant density is omitted. 
}               
\end{table} 
} 
{\narrowtext    
\begin{figure}      
  \begin{center}     
\leavevmode     
        \epsfxsize=4.33cm        
         \epsfbox{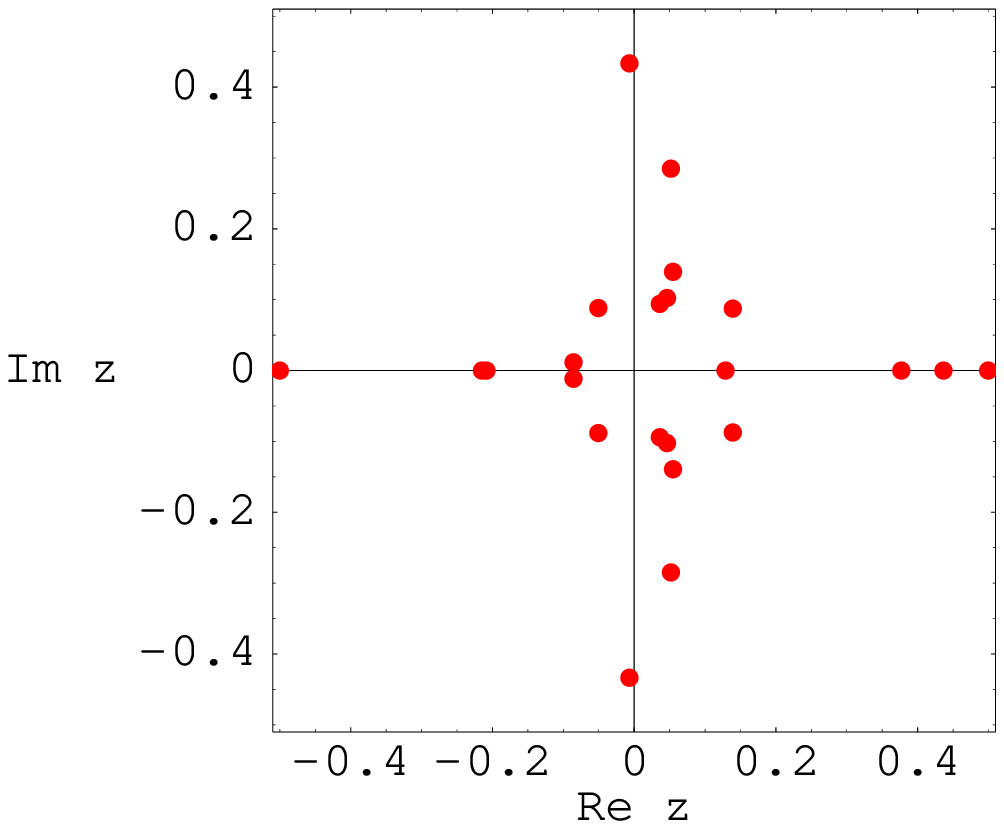}      
\leavevmode     
        \epsfxsize=4.0cm        
         \epsfbox{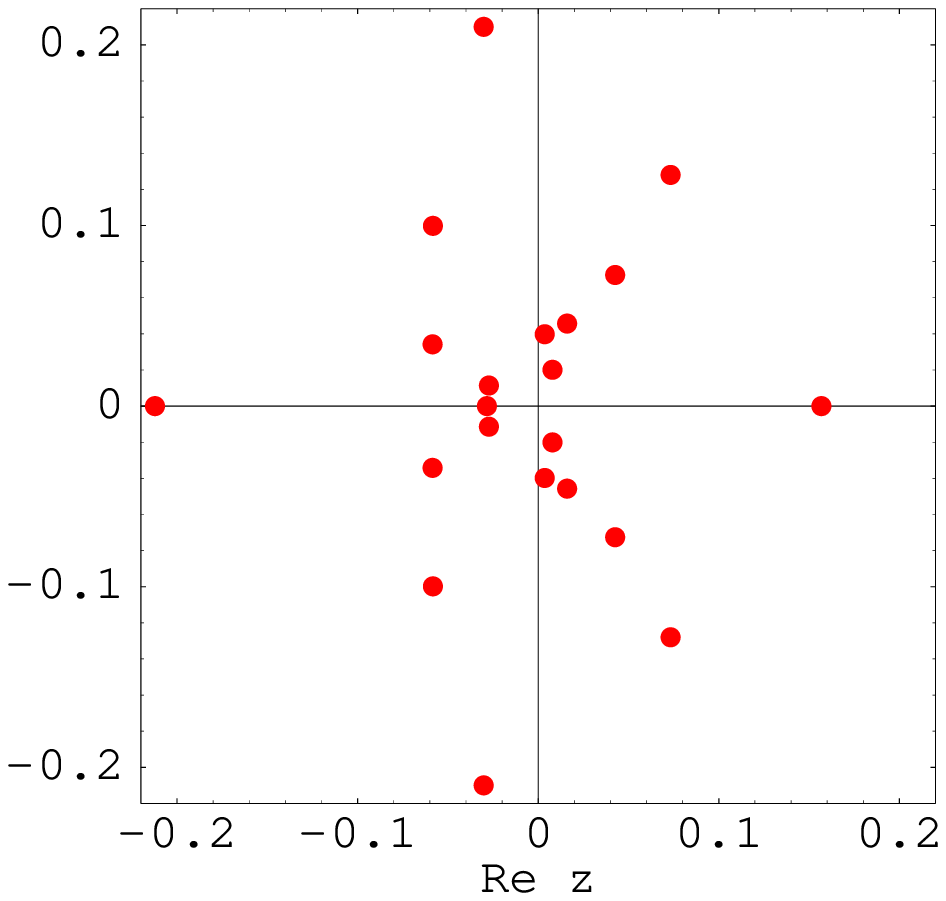}   
\end{center}     
   
	\caption{The configuration, in complex plain, of the Ruelle resonances
of two generic chaotic maps. The right panel 
corresponds the perturbed cat map at k=0.5, and the left panel to the 
standard map at $K=13$. In both cases the coarse graining parameter is 
$s=0.005$. 
}               
\end{figure}        
} 
If we truncate the matrices in (\ref{EVE}) 
to $4\times4$ matrices, the leading Ruelle resonances are the zeros
of the characteristic polynomial: 
\[
\mbox{Det} \left( 
\begin{array}{cccc}  
|T_{5}|^2 &  0 &   
T_{4}^* & (T_{3}^2)^* -z  T_{4}^* \\  
0 &  -z &   
0  &  T_{4}^*  \\  
T_{4} & 0 &   
-z & 0 \\ 
T_{3}^2 -z T_{4} &   
T_{4} &   
0 & |T_{5}|^2 
\end{array} \right) =0. \] 
\noindent
These zeros can be calculated exactly, but, having only truncated matrices, 
it is sensible to evaluate them only to the leading 
orders in $k$. The results are:
\begin{eqnarray} 
z_{1,2}  \simeq \pm \frac{k}{\sqrt{8}}- \frac{k^2}{8}, ~~~ 
z_{3,4} \simeq \pm i \frac{k}{\sqrt{8}}- \frac{k^2}{8}. \label{FR} 
\end{eqnarray}

In Table I we present the numerical (top line) and the variational
results (bottom line) for the Ruelle resonances of the perturbed cat 
map at various values of the perturbation parameter $k$.  
The numerical calculation is performed by projecting the equations (\ref{map}) 
onto a grid of $90\times 90$ sites, and calculating the eigenvalues
at $s=0$ by extrapolation from the interval $0.001\leq s\leq 0.005$.
It has been checked that the results are independent of the grid, 
provided the lattice constant is sufficiently small.
Comparing the results of Eq.~(\ref{FR}) to the exact numerical values, 
shows that the variational approach provides the order 
of magnitude and the correct configuration of the 
leading Ruelle resonances.

To check the generality of this variational approach
we turn now to consider the 
standard map\cite{Chirikov79} as our second example. 
The standard map is defined by 
\begin{eqnarray} 
\begin{array}{l} x_{n+1} = x_n + y_n \\ 
                 y_{n+1}=  y_n +  g(x_{n+1}), 
\end{array} ~~~(\mbox{Mod}~1) \label{Smap} 
\end{eqnarray} 
where  
\[
g(x)= \frac{K}{2 \pi} \sin( 2 \pi x), \]
and $K$ is the stochasticity parameter. The map
is integrable when $K=0$, and it becomes increasingly chaotic
as $K$ is turned on to a large value. The route into chaos
follows the Kolmogorov-Arnold-Moser scenario of breaking 
of resonant tori. Therefore, the standard map
represents a generic mixed system. Yet, for large $K$,
the islands of stability are tiny and have negligible influence
on correlation functions of sufficiently smooth observables. 

Applying the same procedure described above,
we obtain that, as before, Eqs.~(\ref{ZME}) are satisfied, but
the other relevant matrix elements are 
$\langle 1 |  -3 \rangle= \langle 2 |  -2 \rangle= J_2(K)$,
$\langle 2 |  -3 \rangle=J_1^2(K)$, and $\langle 2 |  3 \rangle=J_3^2(K)$, 
where $J_\nu(K)$ is the Bessel function of integer order. Inserting these
results into the truncated eigenvalue equation and calculating 
its zeros we obtain:
\begin{eqnarray}
z_{1,2}= \eta_+(K)\pm \sqrt{\eta_+^2(K)+ J_2(K)}, \nonumber \\
z_{3,4}= \eta_-(K)\pm \sqrt{\eta_-^2(K)- J_2(K)}, \label{zkik}
\end{eqnarray}
where
\[ \eta_\pm(K) = \frac{ J_1^2(K) \pm J_3^2(K)}{2 J_2(K)}.
\]
Thus, to the leading approximation in $1/K$, the first
Ruelle resonances are\cite{comment2}  $\pm |J_2(K)|^{1/2}$
and  $\pm  i |J_2(K)|^{1/2}$. As evident from Eqs.~(\ref{zkik}), $z_{1,2}$
diverge whenever $K$ is a  
zero of $J_2(K)$. At these points our variational approach breaks down, 
but away from them, the results are in agreement with 
the numerical diagonalization, as demonstrated in Table II.

The intriguing feature of the above results is that the configuration
of the leading Ruelle resonances in both examples is similar. 
The four subleading
resonances are located, approximately, at the roots of the equation 
$z^4= \gamma$ where $\gamma$ characterizes 
the stochasticity of the map. 
The more stochastic is the map, the smaller is $\gamma$. In particular,
for the perturbed cat map $\gamma = k^4/64$ while 
for the standard map $\gamma = J_2^2(K)$. It is suggestive 
that this behavior is generic to a wide class of chaotic maps.

In Fig.~2 we depict the positions, in the complex plain, of
the leading 24 resonances of the perturbed cat map and the standard map
with a finite coarse graining, $s=0.005$. 
This figure suggests that the similarity between the classical
spectral properties of the map may extend
 beyond the four subleading resonances.
However, at this stage we do not know how to quantify this similarity. 
{\narrowtext    
\begin{table} 
\vskip 0.7cm
\begin{tabular}{|l|l|l|l|} 
 $K$ & $z_1$ & $z_2$ & $z_{3,4}$   \\ \hline 
10 & 0.577 & $\!-0.526$ & 
$\!-0.064 \!\pm\! 0.521 i$     \\ 
& 0.515& $\!-0.494$ &  $\!-0.003 \!\pm\! 0.505 i$   \\  
\hline
13 & $0.617$  &  $-0.561$ 
&    $-0.002\pm 0.469 i$ \\ 
 & $0.455$ & $-0.478$ & $-0.011\pm 0.466 i$ 
\end{tabular}  
\vspace{0.3cm} 
\caption{The numerical and the variational results for 
the Ruelle resonances of the standard map at $K=10$ and $K=13$.
In each case, the top line is the exact numerical value, 
while the bottom line is the result of the variational calculation (Eqs. 16).  
}               
\end{table} 
} 
\noindent

It is natural to ask what are the implications of the above
results for the spectral statistics of the corresponding quantum maps?
To give a partial answer to this question, we consider the form factor 
which is the Fourier 
transform of the spectral two-point correlation function.
Assuming the map to belong to the orthogonal ensemble,
the semiclassical approximation to the form factor is\cite{AAA}

\begin{equation}
 S(n) \simeq 2 n  \sum_{\alpha =0}^\infty z_\alpha^{n}, \label{FF}
\end{equation}

\noindent
where $n$ denotes an integer time, assumed to be much 
smaller than the Heisenberg time\cite{comment3}, 
and $z_\alpha$ are the Ruelle resonances
of the corresponding classical map. 
The leading resonance, $z_0=1$, is
associated with  the universal result of RMT, while higher
resonances give nonuniversal contributions. As the map becomes 
more chaotic all $|z_\alpha|$ approach zero except for $z_0$, 
therefore the non-universal corrections to RMT become small. 
Below we show that
the configuration of the resonances also plays an important role in
suppressing the magnitude of the nonuniversal contributions.

Let us assume that $z_\alpha \approx 0$ for $\alpha >4$, 
and approximate the 
subleading resonances by $\pm Z$ and $\pm i Z $, where $Z$ is 
a real positive number smaller than unity. 
Substituting these $z_\alpha$-s in (\ref{FF}) one obtains: 
\[ S(n)\approx  2 n + \left\{ \begin{array}{cl}
8 n  Z^n & \mbox{if $\frac{n}{4}$ is an integer} \\
0 & \mbox{otherwise} \end{array} \right. 
\] 
From this formula it follows that the nonuniversal corrections to RMT
appear only in powers of $Z^4$, rather than $Z$. This is a result of
cancellations among the contributions of the Ruelle resonances.
Thus the magnitude of the nonuniversal contribution to the form factor
is dictated both by the absolute values of $z_\alpha$-s, as well as 
by their configuration in the complex plain. It is plausible that
 other chaotic systems exist where the configuration of
resonances lead to an even stronger suppression of nonuniversal contribution,
e.g.~if the subleading resonances  are approximately 
the roots of $z^{2\nu}=\gamma$, where $\nu \geq 3$.
 
To summarize, in this paper we have studied the leading Ruelle resonances of 
two maps representing typical chaotic behaviors: The perturbed cat map which 
exhibits hard chaos, and the standard map which is a mixed system. 
Our analytical and numerical results 
show that, in both cases, the configuration of 
the leading Ruelle, in the complex plain, is similar.
Numerical studies (e.g.~Fig.~2) suggest that  the
similarities in the classical spectrum of chaotic maps 
go beyond the properties of the first four subleading resonances. 
 A comprehensive understanding of the
classical spectral properties of chaotic systems will open the
possibility for understanding the behavior of their quantum counterparts.  
In particular the weak localization mechanisms associated with
quantum interference. In this work we show 
that the configuration of the Ruelle resonances may result in a
suppression of nonuniversal contribution to the form factor.
This mechanism of suppression is different from that of diffusive systems 
in which only the magnitude of the Ruelle resonances is important.

This research was supported by THE ISRAEL SCIENCE FOUNDATION founded by
The Israel Academy of Science and Humanities, and by
Grant No.~9800065 from the USA-Israel   
Binational Science Foundation (BSF).

\end{multicols} 
\end{document}